 \documentclass[10pt,letterpaper,twocolumn]{article} 

\usepackage{ol2}
\usepackage[draft]{hyperref}
\usepackage{amsmath}
\usepackage{graphicx}
\begin{document}
\twocolumn[
\title{Low-loss grating for coupling to a high-finesse cavity}

\author{A. Bunkowski, O. Burmeister, P. Beyersdorf, K. Danzmann, and R. Schnabel}

\address{Max-Planck-Institut f\"ur Gravitationsphysik
(Albert-Einstein-Institut), and Institut f\"ur Atom- und
Molek\"ulphysik, \\Universit\"at Hannover, Callinstr. 38, 30167
Hannover, Germany}

\author{T. Clausnitzer, E.-B. Kley, and A. T\"unnermann}

\address{Institut f\"ur angewandte Physik,
Friedrich-Schiller-Universit\"at Jena, Max-Wien-Platz 1, 07743
Jena, Germany}

\begin{abstract}
A concept for a low loss all-reflective cavity coupler is
experimentally demonstrated at a wavelength of 1064\,nm. A
1450\,nm period dielectric reflection grating with a diffraction
efficiency of 0.58\,\% in the -1st order is used in 2nd order
Littrow configuration as a coupler to form a cavity with a finesse
of 400. The application of such reflective low-loss cavity
couplers in future generations of gravitational-wave detectors as
well as some implementation issues are discussed.
\end{abstract}

]
An international network of first-generation, kilometer-scale,
earth-bound laser-interferometric gravitational-wave (GW)
detectors consisting of LIGO~\cite{LIGO,GWinterferometer04},
GEO\,600~\cite{GWinterferometer04}, TAMA\,300~\cite{TAMA}, and
VIRGO~\cite{VIRGO04} is currently moving from the commissioning
phase to the long-term data-taking operational phase. These
detectors are Michelson interferometers. Power recycling and arm
cavities are two techniques being used to increase the laser power
in the interferometer and hence the detector sensitivity. Both
techniques utilize cavities to which laser light is coupled via a
partially transmitting mirror. For first generation detectors the
light power inside the interferometer will be in the order of
10\,kW at a wavelength of 1064\,nm. To increase the detection
sensitivity even further future GW interferometers will use light
power in the order of MW for which heating effects in the
transmissive elements become an issue. Power absorption in the
substrates leads to thermal lensing and also to deformation of the
optical surface. These distortions will limit the circulating
power below the level which is necessary to optimize quantum
noise. To reduce thermal noise cryogenic techniques for the main
optics are very likely to be used in third generation GW
detectors. Absorbed heat in the substrates will worsen the cooling
efforts of the optical elements. To avoid heating in the substrate
reflective grating beam splitters can be used instead of partially
transmissive mirrors and beam splitters~\cite{Drever96}. An
additional advantage of all reflective optics within  GW detectors
is the elimination of the constraint that the substrate materials
be optically transparent. Considering opaque substrate materials
with superior mechanical properties allows lowering the thermal
noise in the detector.

\begin{figure}[h]\centerline{\scalebox{0.7}{\includegraphics{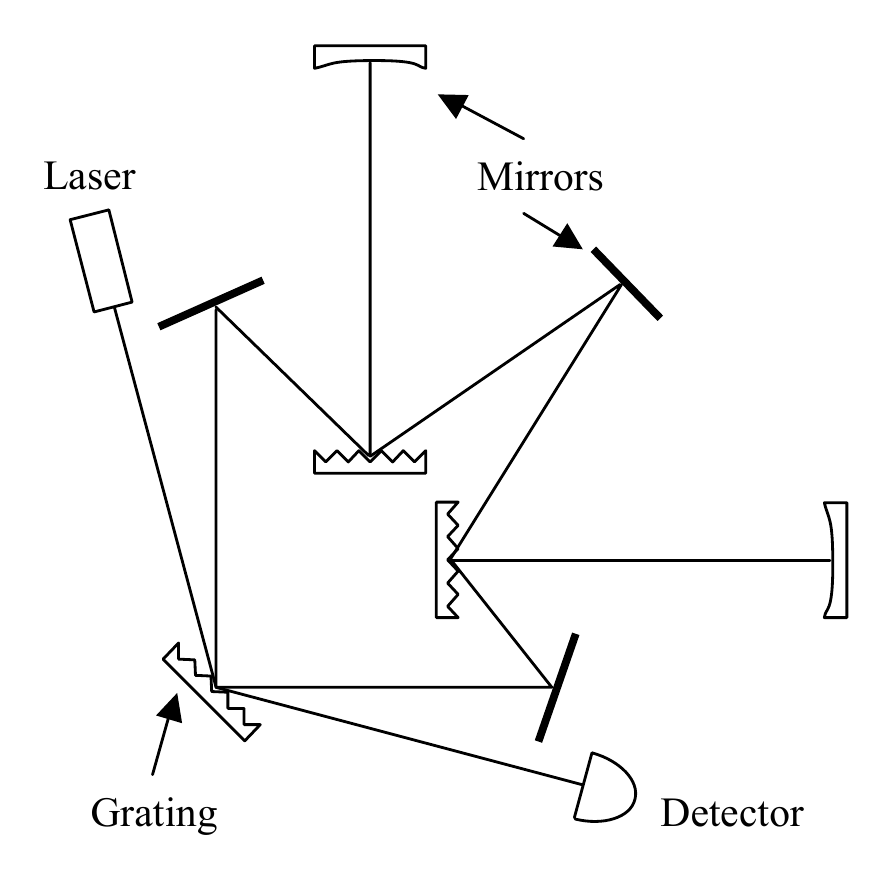}}}
  \caption{Sketch of the interferometer setup proposed by
  Drever.}
   \label{dreversetup}
\end{figure}

In proof-of-principle experiments Sun and Byer~\cite{Sun97}
demonstrated a Michelson and a Sagnac interferometers based on
all-reflective elements. They also demonstrated a Fabry-Perot
coupler concept that is based on high diffraction efficiency
gratings in 1st order Littrow configuration. Drever pointed out
that low diffraction efficiency gratings could also be used as
cavity couplers and argued that the overall losses should be lower
than in high diffraction efficiency elements. An all-reflective
interferometer configuration which avoids the use of a 50/50
beamsplitter and uses low diffraction efficiency gratings and
mirrors as the only major optical elements as schematically
depicted in figure~\ref{dreversetup} was proposed~\cite{Drever96}.
But to our knowledge no experimental realization of
interferometers utilizing low diffraction gratings has been
reported so far.

In this letter we report on the design of a low-loss diffraction
grating with a diffraction efficiency of less than 1\,\% and on
the experimental realization of a high-finesse linear cavity using
the grating as a coupler.

\begin{figure}[h]\centerline{\scalebox{0.9}{\includegraphics{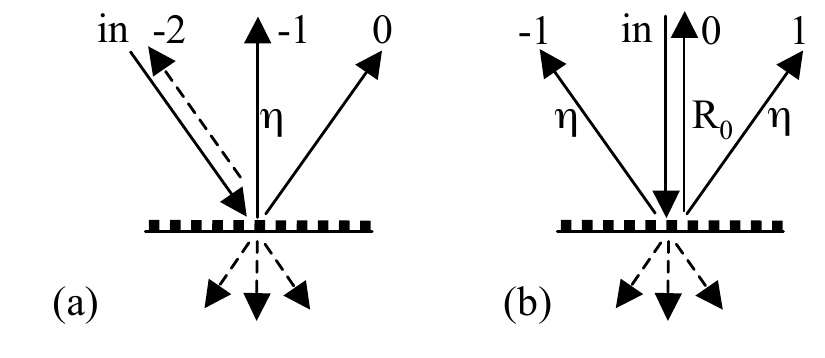}}}
  \caption{Reflected and transmitted orders from a reflection grating in
  2nd order Littrow configuration; (a) Incident beam at the Littrow angle
  (b) Normal incidence.}
  \label{scheme}
\end{figure}

For a laser beam of wavelength $\lambda$ incident on a reflection
grating, the output angle $\Theta_m$ of the $m$th diffracted order
is given by the well known grating equation
\begin{equation}
d(\sin\Theta_m-\sin\Theta_{in})=m\lambda,
\end{equation}
where $\Theta_{in}$ is the incidence angle and $d$ is the grating
period. If used in 1st order Littrow configuration
$(\sin\Theta_{in}= \lambda /2d)$ a reflection grating can be used
as a cavity coupler\cite{Sun97}; the \emph{reflected} (0th order)
beam is used to couple into the resonator. The finesse of such a
resonator is limited by the maximum diffraction efficiency of the
first order of the grating. However if used in 2nd order Littrow
configuration, as shown in figure~\ref{scheme} the
\emph{diffracted} (-1st) order is used to couple to the cavity.
Then the maximum reflectivity R$_0$ of the grating under normal
incidence is the limiting factor for the finesse of the cavity.

Standard coating techniques can routinely produce multi-layers
with a reflectivity greater than 99.98\,\%. This value is to be
compared with routinely produced maximum diffraction efficiencies
of about 95\,\% \cite{Clausnitzer03}. To our knowledge the highest
value ever reported so far is also not greater than
99\,\%\cite{Britten02}. Therefore the 2nd order Littrow
configuration is the appropriate choice for efficient low-loss
coupling to a linear resonator.

Every diffraction order that is allowed by the grating equation
will contain some light power. To reduce overall losses in the
device the grating period $d$ should be chosen in a way that only
the diffraction orders that are going to be used are allowed by
the grating equation. Only the $\pm1$ orders for normal incidence
are needed in our case which suggests
\begin{equation}
\lambda<d<2\lambda.
\end{equation}

\begin{figure}[h]\centerline{\scalebox{0.4}{\includegraphics{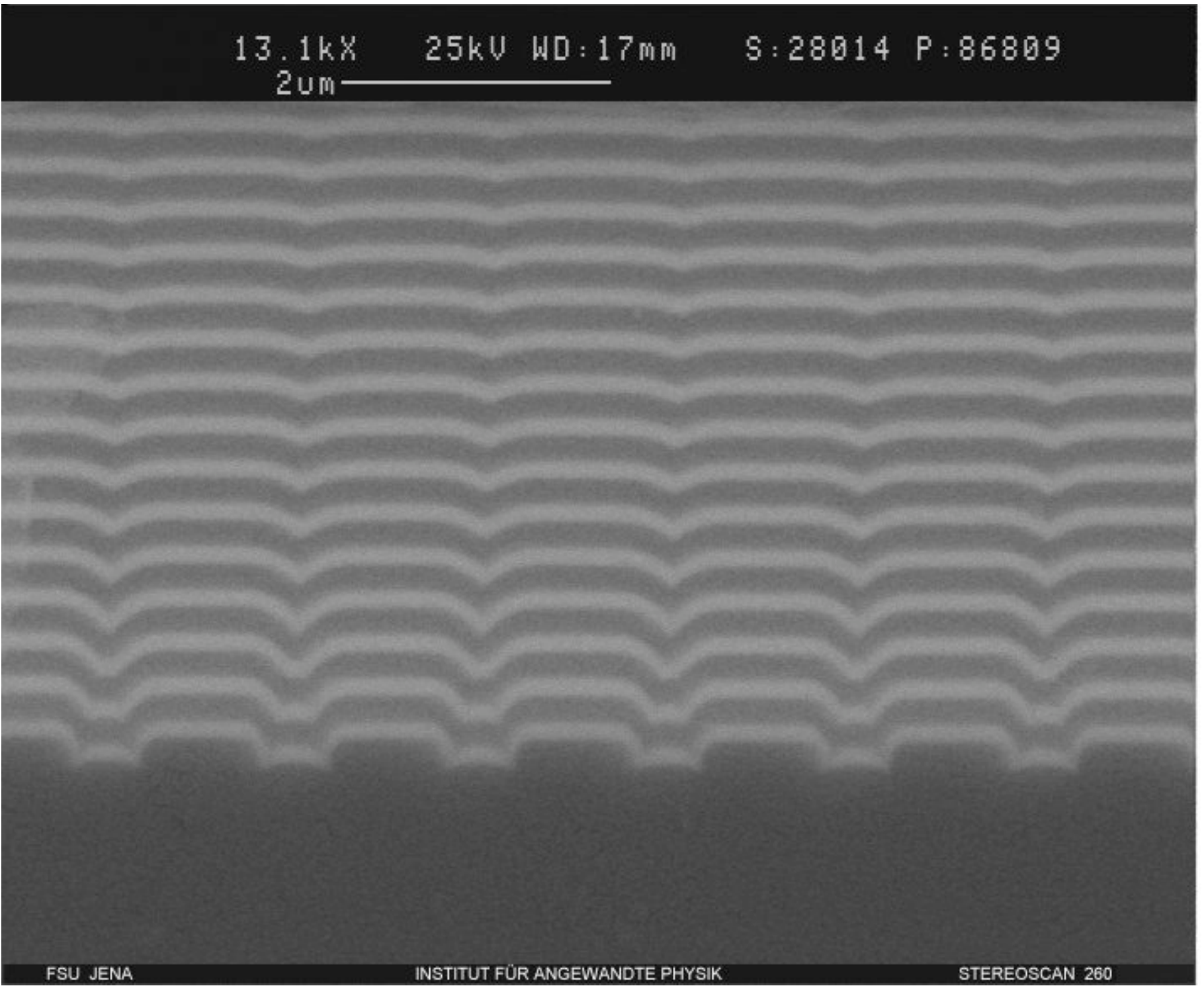}}}
  \caption{Cross section of an overcoated binary grating (SEM-image).}
   \label{pic}
\end{figure}

A common way of manufacturing high efficiency dielectric
reflection gratings is to etch a periodic structure into the top
layer of a dielectric multilayer stack as it is done for the
gratings used in high-power chirped-pulse
amplification~\cite{Shore96}. For the low diffraction efficiency
grating needed for our application we used a different approach.
We first etched the grating into a substrate and then overcoated
it so that the dielectric layers effectively form a volume grating
as can be seen in figure \ref{pic}. A  shallow binary structure
with a depth of 40-50\,nm, a ridge width of 840\,nm and a period
of $d=1450\,$nm was generated by electron beam lithography and
reactive ion beam etching on top of a fused silica substrate. The
applied multilayer stack was composed of 32 alternating layers of
silica (SiO$_2$) and tantalum pentoxide (Ta$_2$O$_5$). The
diffraction efficiency for 1064\,nm light with a polarization
plane parallel to the grating grooves and perpendicular to the
plane of incidence ($s$-polarization) was measured to be
$\eta=(0.58\pm0.04)\,\%$.

\begin{figure}[h]\centerline{\scalebox{0.85}{\includegraphics{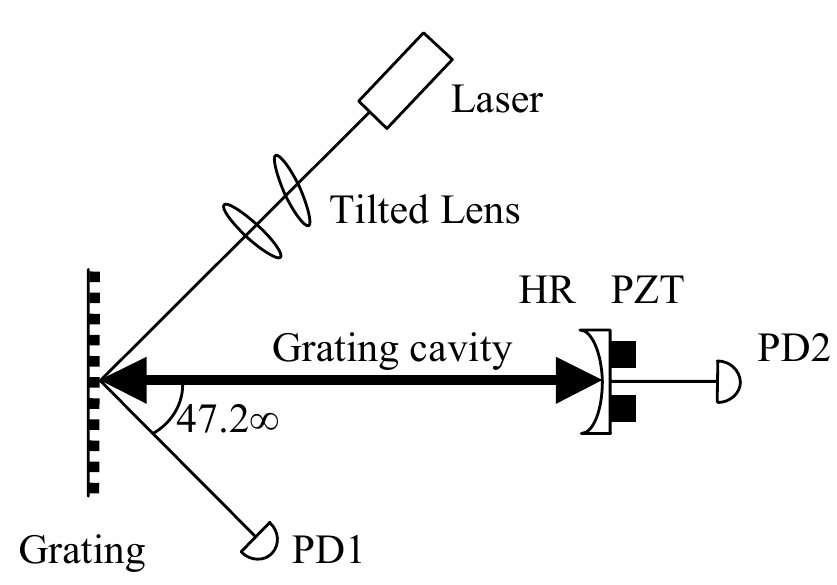}}}
  \caption{Experimental setup for the demonstrated grating Fabry-Perot cavity: HR,
  high reflector; PZT, piezoelectric transducer; PD, photo diode.}
   \label{setup}
\end{figure}

Figure~\ref{setup} shows the experimental setup for the
all-reflective Fabry-Perot cavity. A high reflective (HR) mirror
with a radius of curvature of 1.5\,m mounted on a piezoelectric
transducer (PZT) to allow for cavity length control was placed
parallel to the grating surface at a distance of 43\,cm. An
$s$-polarized beam of 50\,mW from a 1.3\,W, 1064\,nm diode pumped
Nd:YAG laser was used. The angle of incidence corresponded to
second order Littrow configuration $\Theta_{in}=
\arcsin(\lambda/d)\approx 47.2\,^{\circ}.$ The circulating and
reflected power from  the cavity were observed using the leakage
from the high reflector and from the 0th order of the grating
respectively.

High circulating powers inside the cavity also demands good mode
matching of input beam and cavity mode. Note that our grating
couples modes at different angles of incident which changes the
ratio of horizontal an vertical mode widths. The following
relation holds for the horizontal width $w_{h}$ of the beam
\begin{equation}
    w_{in,h}/w_{m,h}=
    \cos \Theta_{in}/\cos \Theta_m,
\end{equation}
where $in$ and $m$ refer to the incoming and the diffracted beam
respectively. For our setup an input beam with an elliptical
profile having a horizontal width of 1.47 times $(w_{h}/w_{v}=
\cos\Theta_{m}/\cos\Theta_{in}\approx 1.47)$ the vertical width
produced the desired round beam profile in the diffracted beam.
The profile is generated by a pair of lenses from which one lens
was tilted horizontally to have different focal lengths for the
$v$- and $h$-directions.

\begin{figure}[h]\centerline{\scalebox{0.55}{\includegraphics{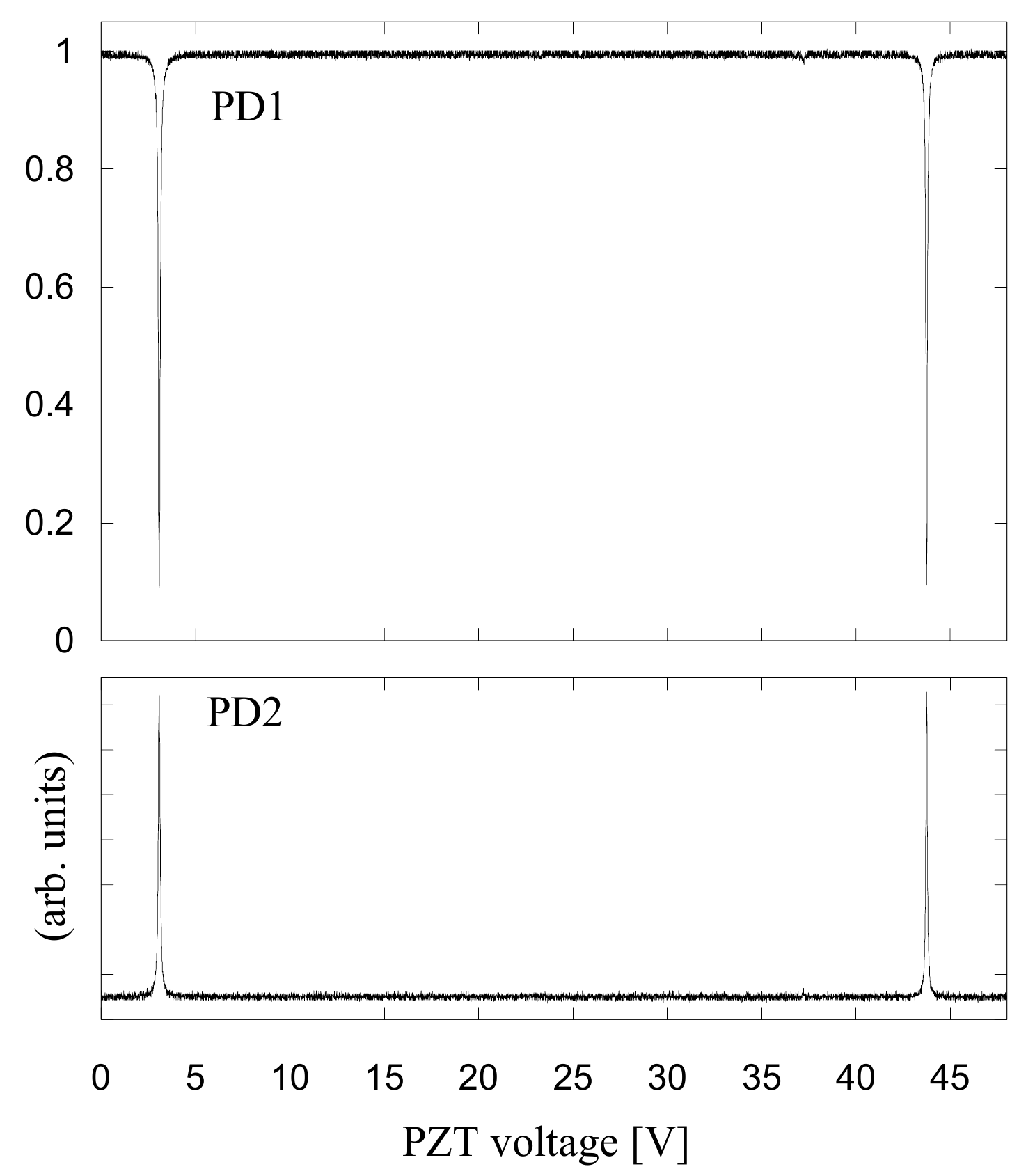}}}
  \caption{Measured signals of the two photodiodes according to
  figure \ref{setup} versus  PZT ramp voltage over one free spectral range.
  PD1, reflected power normalized to the power of the incident beam; PD2,
  circulating power in the cavity.}
   \label{pdsignals}
\end{figure}

Figure~\ref{pdsignals} shows the transmission and reflection
interference fringes for the cavity with a measured cavity finesse
of $400\pm20$. More than 99\,\% of the power was measured to be in
the TEM$_{00}$ mode indicating excellent mode matching. With the
measured value of the finesse and the known reflectance of the
HR-mirror one can calculate the overall losses $A$ which are
defined by $A= 1-R_0-2\eta$, where $R_0$ is the the reflectance
for normal incidence. We find $A=(0.38\pm0.2)\,\%$. Losses are due
to transmitted orders and scattered light.
The partially transmissive mirrors conventionally used for
coupling to a linear Fabry-Perot cavity can be considered as
2-port devices with simple phase relations between the two ports.
The input coupler introduced here, however, is a 3-port device.
Light entering one port will always couple to all three ports. The
phase relations of the light in the three different ports are more
complex than in a 2-port device. They depend on the diffraction
efficiencies for the different orders  and can be calculated with
a scattering matrix formalism~\cite{Siegman}. Due to the
additional port new GW detector topologies can be accomplished.
The scheme shown in figure \ref{dreversetup} for example uses two
linear grating cavities in the arms. On resonance these cavities
will retro-reflect most of the power incident onto the grating
while the signal sidebands generated in the arms will be split
equally between the two output ports of the cavity. Therefore
power and signal are taking different paths in the interferometer.
A detailed analysis of the phases of the three ports of the
coupler as well as their effects on the properties of the proposed
interferometer is in preparation and will be presented in a
forthcoming paper.

In conclusion we have experimentally demonstrated that a low
efficiency grating can be used as a cavity input coupler with low
loss. A cavity with a finesse of 400 was constructed far exceeding
the best finesse values for an all-reflective cavity that had been
previously reported. In a next step we will optimize design and
manufacturing process of the gratings to produce gratings with
even lower diffraction efficiency and overall losses. These
gratings will have high potential to be implemented in future GW
detector configurations.

This research was supported by the Deutsche Forschungsgemeinschaft
within the Sonderforschungsbereich TR7.

\end{document}